\documentclass[aps, prd, nofootinbib, amsmath, twocolumn, preprintnumbers]{revtex4-1}
\usepackage{amssymb,esvect,amsmath,graphicx,latexsym,amsthm,slashed,eso-pic,hyperref}
 \DeclareMathOperator{\kev}{keV} \DeclareMathOperator{\mev}{MeV} \DeclareMathOperator{\gev}{GeV}  \DeclareMathOperator{\cm}{cm}         \DeclareMathOperator{\few}{few} 
        \newcommand{\cO}{{\cal O}}

\newcommand{\pL}{\left(} \newcommand{\pR}{\right)}      
\newcommand{\beq}{\begin{equation}} \newcommand{\eeq}{\end{equation}}
\newcommand{\bea}{\begin{eqnarray}} \newcommand{\eea}{\end{eqnarray}}
\newcommand{\alg}[1]{\begin{align} \begin{split} #1 \end{split}  \end{align}}

\newcommand{\Eq}[1]{Eq.~(\ref{#1})} \newcommand{\Eqs}[2]{Eqs.~(\ref{#1}) and (\ref{#2})} 
\newcommand{\Sec}[1]{Sec.~\ref{#1}} \newcommand{\Secs}[2]{Secs.~\ref{#1} and \ref{#2}} 
\newcommand{\Fig}[1]{Fig.~\ref{#1}}

\newcommand{\be}{\begin{eqnarray}}
\newcommand{\ee}{\end{eqnarray}}

\begin{document}

\preprint{FERMILAB-PUB-19-358-A} 

\title{Implications of BBN Bounds for Cosmic Ray Upscattered Dark Matter}

\author{Gordan Krnjaic}
\thanks{ORCID: http://orcid.org/0000-0001-7420-9577}
\author{Samuel D.~McDermott}
\thanks{ORCID: http://orcid.org/0000-0001-5513-1938}

\affiliation{Fermi National Accelerator Laboratory, Theoretical Astrophysics Group, Batavia, IL, USA}

\begin{abstract}
We consider the Big Bang Nucleosynthesis (BBN) bounds on light dark matter whose cross section off nucleons is sufficiently large to enable
acceleration by scattering off of cosmic rays in the local galaxy. Such accelerated DM could then deposit energy in terrestrial detectors. Since this signal involves DM of mass $\sim \kev - 100\mev$ {\it and} requires large cross sections $\gtrsim 10^{-31}$ cm$^2$ in a relativistic kinematic regime, we find that 
the DM population in this scenario is generically equilibrated with Standard Model particles in the early universe. For sufficiently
low DM masses $\lesssim 10$ MeV, corresponding to the bulk of the favored region of many cosmic-ray upscattering
studies, this equilibrated DM population adds an additional component to the relativistic energy density 
around $T \sim$ few MeV and thereby spoils the successful predictions of BBN. In the remaining $\sim 10-100 \mev$ mass range, the large couplings required in this scenario are either currently excluded or within reach of current or future accelerator-based searches.
\end{abstract}

\maketitle

\section{Introduction}
\label{intro}

Although the gravitational evidence for dark matter (DM) on large scales is overwhelming, its microscopic properties are
 unknown and consistent with a wide array of theoretical possibilities
  \cite{Bertone:2016nfn}. In light of null results from direct-detection searches for weak-scale DM, 
 much attention has recently focused on the $\lesssim$ GeV mass range
which is inaccessible to these conventional probes (see \cite{Battaglieri:2017aum} for a review).

Traditional nuclear-recoil direct detection experiments are insensitive to 
sub-GeV DM because the local virial velocity $v \sim 10^{-3} c$ is insufficient for light DM to impart 
 nuclear recoil energies above observable experimental thresholds.
 However,  it has recently been shown that for light DM in the keV-MeV mass range, 
 some fraction of the halo population will be {\it upscattered} by cosmic ray (CR) interactions, which accelerate
 the light DM to quasi-relativistic speeds that can impart observable energy to  terrestrial nuclear targets 
 \cite{Cappiello:2018hsu, Bringmann:2018cvk, Cappiello:2019qsw}.
 We refer to this scenario as Cosmic Ray Upscattered Dark matter (CRUD). 
 
 In order for CRUD to occur with a detectable rate for typical
 experimental exposures, the DM-nucleon cross section must be quite large $\gtrsim 10^{-31} {\rm\, cm^2}$ 
 to overcome the low probability of being upscattered by a CR \cite{Cappiello:2018hsu}.
Large scattering cross sections for dark matter particles at and above the GeV scale have been considered before,
and are subject to a wide variety of limits \cite{Starkman:1990nj, Wandelt:2000ad, Mack:2007xj}. Extending these investigations to the 
sub-GeV-mass regime is novel, and as we discuss below, subject to qualitatively different constraints.

We begin by noting that Refs. \cite{Cappiello:2018hsu, Bringmann:2018cvk, Cappiello:2019qsw}
 assume the cross section to be constant and identical in both the CR upscattering and conventional direct detection contexts. Although we will revisit this assumption (and later argue that it is unphysical), for now we temporarily adopt this constant cross-section ansatz to heuristically argue that this scenario faces stringent cosmological bounds.
A key observation that we will apply throughout this work is that  the cosmic ray interactions required to upscatter DM necessarily yield {\it  highly 
 relativistic} DM particles. For this reason, it is self-consistent to extrapolate this constant cross section to the early universe
 independent of other DM properties (e.g. spin, chirality, or Lorentz structure of interactions).

By crossing symmetry, the $ \chi q \to  \chi q$ scattering cross section
is related to the  $\bar q q \leftrightarrow \bar \chi \chi$ creation/annihilation cross section. In  
the relativistic regime, these differ only by order-one factors,  
so it is interesting to ask: for what cross sections does the  DM population come into thermal equilibrium
 with the Standard Model (SM)?
To get a rough sense of the answer, which we will refine below, we compare the rate of dark matter creation at the temperature of the
 QCD phase transition $T \simeq \Lambda_{\rm QCD}$ to the Hubble rate at that time.
 With this crude estimate, the critical DM equilibration cross section $\sigma_{\chi q}^{\rm (eq)}$ is
 \alg{ \label{schematic}
 n_q \sigma_{\chi q}^{\rm(eq)} &\simeq \Lambda_{\rm QCD}^3 \sigma_{\chi q}^{\rm(eq)} \simeq \sqrt{g_\star} \Lambda_{\rm QCD}^2/M_{\rm Pl} 
\\ &\implies \sigma_{\chi q}^{\rm(eq)} \simeq \frac{\sqrt{g_\star}}{ \Lambda_{\rm QCD} M_{\rm Pl} } \simeq 10^{-46}\cm^2,
}
where $g_\star \sim 60$ is the number of light species just before this transition.
Thus, for all scattering cross sections of interest in Refs.
\cite{Cappiello:2018hsu, Bringmann:2018cvk, Cappiello:2019qsw} it is clear that
the DM thermalizes with the SM in the early universe.
Since the predictions of big bang nucleosynthesis (BBN) work well with Standard Model particle content only,
the presence of one or more new thermalized degrees of freedom is prohibited, as we explain in more detail below.
This argument excludes thermalized DM below  $m_\chi \lesssim 5\mev$, which covers much of the $\sim \kev - \,100\mev$ parameter space over which
CRUD can yield observable rates. The remaining $\sim$ 5$-$100 MeV range is likely within reach of existing and future 
accelerator searches for light DM.

However, contrary to the assumption made in \cite{Cappiello:2018hsu, Bringmann:2018cvk, Cappiello:2019qsw} we will show that the cross section for scattering with a cosmic ray is not equal (or comparable) to the ``momentum-independent'' cross section canonically probed by direct detection experiments in the broad, representative class of interactions studied below. Such cross sections cease being momentum-independent when the particle momenta become large compared to their masses, as is the case in cosmic ray collisions. This issue was anticipated by \cite{Cappiello:2018hsu, Cappiello:2019qsw}, and in the remainder of this paper we explore the consequences of this energy dependence.

This paper is organized as follows. \Sec{BBN} presents the BBN-only bound on light, thermalized particle species. \Sec{contact} generalizes the argument in \Eq{schematic} to a contact interaction, which we find to be excluded for most of the CRUD parameter space. \Sec{mediators} extends this argument further to include interactions mediated by a light (or massless) particle.  \Sec{other} excludes other exotic possibilities, \Sec{lab} discusses laboratory bounds for the BBN-safe mass range, and  \Sec{conclusion}
offers some concluding remarks.

\section{BBN and Light Thermal DM}
\label{BBN}
Standard BBN theory successfully accounts for the light-element yields observed in
the universe today (see \cite{Tanabashi:2018oca,Cyburt:2015mya} for reviews). These predictions depend only on two cosmological parameters: 
the baryon-to-photon ratio $\eta_b = n_b/n_\gamma$, which sets the nucleon density, and the Hubble expansion rate $H_{\rm BBN}$, which governs the duration of BBN. Since $\eta_b = 6.13 \pm 0.03 \times 10^{-10}$ is precisely known from independent measurements of CMB anisotropies \cite{Cyburt:2015mya, Aghanim:2018eyx} and is difficult to change with new physics, the expansion rate
can be used to constrain light thermalized particles. 

The effect of a new species with a thermal abundance is to increase the expansion rate due to its 
contribution to the Hubble parameter, since $H^2 \propto \rho_{\rm tot}$. Assuming there is not a large 
DM chemical potential, any species $\chi$ that is in thermal and chemical equilibrium with the SM bath will contribute to 
$\rho_{\rm tot}$ in a way that is determined strictly by $m_\chi$ and $T$.
By rescaling the value of D/H in the large-$m_\chi$ limit of Fig.~7 of \cite{Nollett:2014lwa}
to values that have been obtained using updated nuclear rates with lower uncertainties \cite{Coc:2015bhi, Berlin:2019pbq},
and comparing these to the recent determination of the D/H abundance in high redshift quasars
\cite{Zavarygin:2018dbk}, we find the limits  
\be
\label{massbound}
m_\chi > 
\begin{cases}
 0.9 \, {\rm  MeV}~  {\rm Real ~scalar}\\ 
  5.3 \, {\rm  MeV} ~{\rm Complex~ scalar} \\
   5.0 \, {\rm  MeV} ~{\rm Majorana ~fermion}\\  
 7.8 \,{\rm  MeV}~{\rm Dirac ~fermion} 
\end{cases}~.
 \ee
This argument relies only on the assumption that 
 $\chi$ reached equilibrium with the SM before BBN and does not depend on the details of its freeze out, the nature of its SM coupling, or whether its population is particle/antiparticle symmetric.  Indeed, the argument applies to particles $\chi$ that constitute an arbitrarily small fraction of the present-day DM density, because their late-time abundance is sensitive to the freeze out of $\chi$, whereas the BBN $N_{\rm eff}$ bound applies to the $\chi$ density {\it before} freeze out (while it is still in equilibrium).
 In all variations of these scenarios, there exists a large, thermal
 number density which increases $H$ at BBN and thereby spoils the successful 
 light-element predictions. 

 This BBN-only bound is qualitatively different from the stronger, but more model dependent, bound on $\Delta N_{\rm eff}$ derived from CMB  temperature anisotropies. Since the CMB bound is sensitive to $T_\nu/T_\gamma$, this ratio can in principle be modified to compensate for the effect of a new thermalized species. By contrast, the BBN bound depends only on the total expansion rate during this epoch,
  so all new relativistic species increase this value and thereby modify light-element yields. 

 For the remainder of this paper, we will apply this argument to diverse representative   
interactions and show that any cross-section sufficiently large to observe CRUD 
necessarily implies that the $\chi$ is thermalized in the early universe.

\section{Contact Interactions}
\label{contact}

\begin{figure*}[t!]
\includegraphics[width=0.78\textwidth]{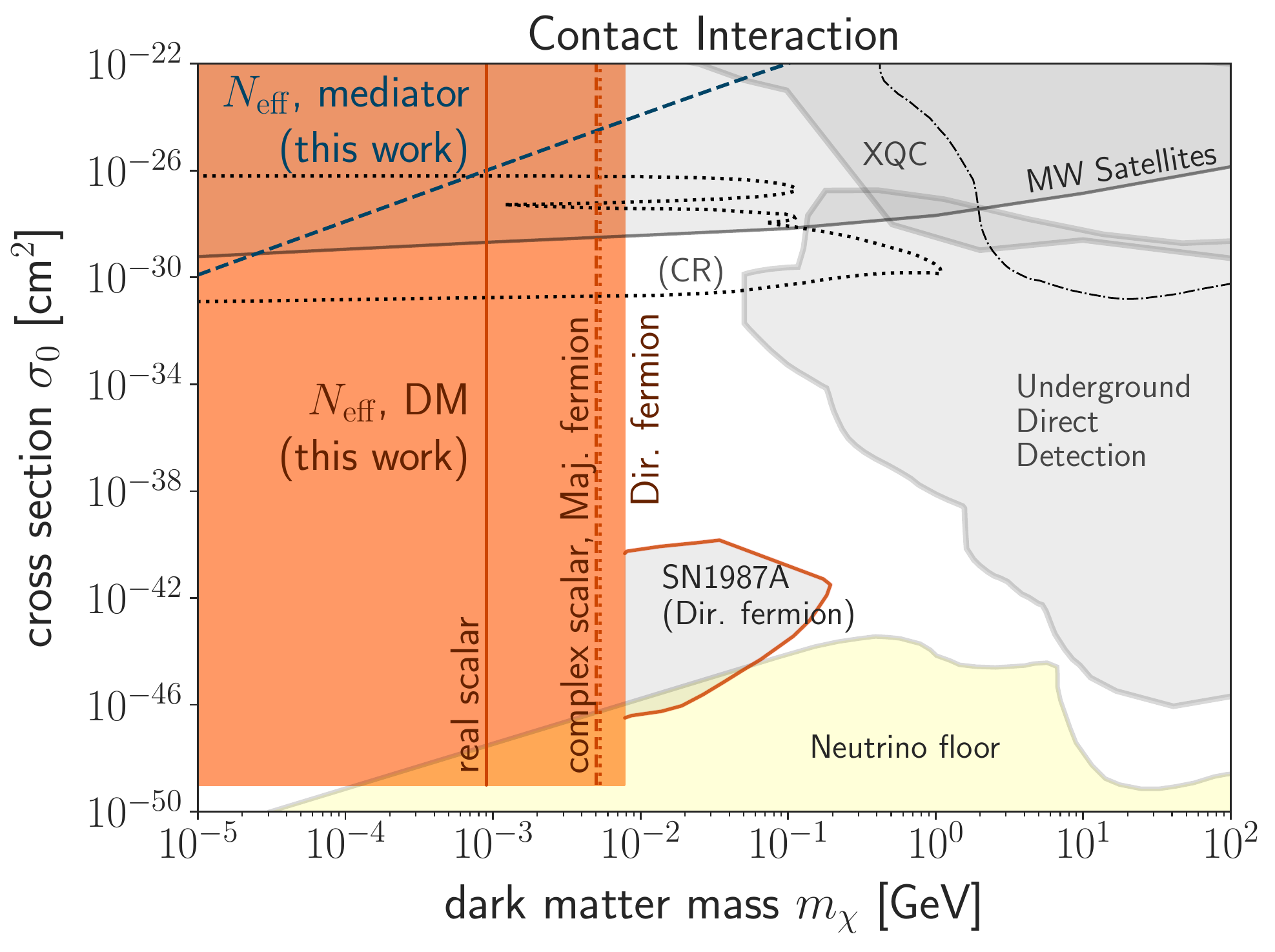}
\caption{
Limits on the direct-detection cross section for the contact interaction in \Eq{Lint}. The red shaded region is  
the Dirac fermion BBN $N_{\rm eff}$ bound from \Eq{massbound} for $\sigma_0 \gtrsim 10^{-49}$ cm$^2$, which saturates 
the inequality in \Eq{hubble-criterion}. The bounds from \Eq{massbound} for a real scalar, a Majorana fermion, and a complex scalar are shown as solid, dashed, and dotted lines, respectively. Although 
the underground direct detection \cite{Akerib:2016vxi, Cui:2017nnn, Aprile:2018dbl, Agnes:2018ves, Abdelhameed:2019hmk, Liu:2019kzq, Davis:2017noy, Kavanagh:2017cru, Emken:2018run,Hooper:2018bfw}, XQC \cite{Wandelt:2000ad, Erickcek:2007jv,Mahdawi:2018euy} and MW Satellite \cite{Nadler:2019zrb} limits assume that the cross section is spin- and momentum-independent in the non-relativistic limit, the $N_{\rm eff}$ exclusion regions 
are valid with or without this assumption. Above the dashed blue line labelled ``$N_{\rm eff}$, mediator'', 
$\sigma_0$ is so large that it requires a light $\leq10\mev$ mediator as in \Eq{mediator-requirement} whose presence 
also spoils BBN. The dashed (CR) contours are the CRUD 
regions from Refs.~\cite{Bringmann:2018cvk, Cappiello:2019qsw}, which
 assume a cross section that is constant and equal for CR scattering and non-relativistic direct detection (see \Secs{intro}{contact} for discussion).
}
\label{fig:money}
\end{figure*}

In \Sec{intro}, we showed that an exactly constant DM-SM cross section large enough to realize CRUD 
would thermalize the DM in the early universe.
In \Sec{BBN}, we argued that this could be problematic for $m_\chi \lesssim \few \mev$. However, as noted
in \cite{Cappiello:2018hsu,Cappiello:2019qsw, Dent:2019krz} and in \Sec{intro}, a constant cross section across all energy scales is physically implausible.
For the rest of this paper, we will 
instead make a series of more realistic assumptions to make the BBN bound more meaningful. 
In this section, we focus on contact interactions between DM and SM quarks.

In a cosmic ray upscattering event, the typical CM energy for a low-velocity DM target is of order  
\be \label{ecr-gev}
E_{\rm CM} \simeq \sqrt{2 E_{\rm CR} m_\chi} = {\rm GeV}  \sqrt{      \frac{E_{\rm CR}}{500\gev} ~\frac{m_\chi}\mev  }~,~~
\ee
assuming that $E_{\rm CR} \gg m_\chi, m_p$.
If the DM-CR interaction is well described by a contact operator, then this contact description
of DM-SM interactions must be valid at GeV-scale energies. Thus, it is a contact interaction for temperatures
of order  $T \lesssim$ GeV, which includes both the QCD phase transition at $T \sim 200$ MeV and the BBN 
epoch at $T \sim$ MeV. Importantly, we show here that the required CRUD coupling to nucleons at late 
times predicts a large {\it annihilation} cross section for $\bar q q \to \bar \chi \chi$ prior to the QCD 
phase transition at $T\sim \Lambda_{\rm QCD} \sim 200$ MeV. This process thermalizes the DM with the SM in the early universe. 

Consider the contact interaction between a light DM candidate $\chi$ and SM quarks
\be
\label{Lint}
{\cal L}_{\rm int} = G_{\chi q} ( \bar \chi \gamma_\mu  \chi)(\bar q \gamma^\mu q) ,~~ T \gg \Lambda_{\rm QCD}.
\ee
The coefficient $G_{\chi q}$ in \Eq{Lint} has mass-dimension $-2$, so this operator is non-renormalizable and arises from integrating out a heavy mediator particle whose mass exceeds the $E_{\rm CM}$ given in \Eq{ecr-gev}.
For temperatures below the QCD phase transition, the interaction in \Eq{Lint} becomes
\be \label{LintN}
{\cal L}_{\rm int} \to G_{\chi N} ( \bar \chi \gamma_\mu  \chi)(\bar N \gamma^\mu N) ,~~ T \ll \Lambda_{\rm QCD},
\ee
where $G_{\chi N}$ is the effective coupling to the nucleons $N=n,p$. This coupling satisfies $G_{\chi N} = \sum_q G_{\chi q}$ for a sum over all valence quarks inside the nucleon.
Although  we have chosen a vector Lorentz structure here for simplicity, both CRUD and early universe DM annihilation are relativistic processes, so our approach is without loss of essential generality; the rates for other Lorentz structures will differ only by factors $\sim \cO(1)$ in the relativistic limit.

The accessible parameter space for CRUD is conventionally presented in terms of the 
{\it non-relativistic} direct detection cross-section $\sigma_0$, where
\be
\label{sigma-dd}
\sigma_{0} \equiv \frac{ G^2_{\chi N} \mu_{\chi N}^2 }{\pi} ~, ~~ \mu_{\chi N} \equiv \frac{m_\chi m_N}{m_\chi + m_N}~~.
\ee
We emphasize that this non-relativistic scattering cross section is appropriate to use for the cold cosmological dark matter at the CMB epoch and in the Milky Way today, but it is not appropriate for describing the collision of a relativistic cosmic ray with a much lighter $\chi$ (nor for the collision of the relativistic outgoing $\chi$ with a direct detection apparatus).
Neglecting quark and DM masses, the {\it relativistic} scattering cross-section (appropriate for upscattering or for relativistic direct detection) and the $\bar q q \leftrightarrow \bar \chi \chi$ annihilation cross section are, respectively,
\be
\label{sigma-rel}
\sigma_R(s) =   \frac s6 \frac{ \sigma_{0}}{\mu_{\chi N}^2}, \qquad \sigma_{\rm ann}(s) =    \frac{     s      }{12} \frac{ \sigma_{0}}{\mu_{\chi N}^2}~,
\ee
where $s = E_{\rm CM}^2$ is the Mandelstam variable and $G_{\chi q}^{-1}\!\gg \! s$ is true in the contact-interaction regime by assumption. Given this discussion it is clear that, even for a contact-interaction Lagrangian, the cross section does not equal the same constant for non-relativistic direct detection and relativistic CR-DM scattering as assumed in \cite{Cappiello:2018hsu, Bringmann:2018cvk, Cappiello:2019qsw}.

Since CRUD requires a DM interaction with light quarks, we conservatively ask: was the DM in chemical equilibrium with the SM at $T\sim \Lambda_{\rm QCD}$ when the universe 
 contained thermal densities of quarks and anti-quarks? 
The order of magnitude criterion for thermalization is that the thermally averaged quark-antiquark annihilation rate  $\Gamma_{\bar q q \to \bar \chi \chi} = n_q \langle  \sigma v \rangle_{\rm ann}$ is more rapid than 
Hubble expansion prior to the QCD phase transition. The thermally averaged annihilation cross section is
\be
\langle  \sigma v \rangle_{\rm ann} =  \frac{1}{N}  \int_{4m_\chi^2}^\infty ds \, \sigma_{\rm ann}(s)\sqrt{s} (s-4m_\chi^2) K_{1} \!\! \left(\frac{\sqrt{s}}{T}\right)\! ,~~~
\ee 
where $N = 8 m_\chi^4 T K^2_2(m_\chi/T)$ and $K_n$ is a  modified Bessel function of the second kind \cite{Gondolo:1990dk}. To assess the value of $\sigma_{0}$ for which
thermal equilibrium is attained, we divide $\Gamma_{\bar q q \to \bar \chi \chi}$ by $H(T_{\rm QCD})$:
\be
\label{hubble-criterion}
\frac{\Gamma_{\bar q q \to \bar \chi \chi} }{H(T_{\rm QCD})} \simeq \frac{   \langle  \sigma v \rangle_{\rm ann}  M_{\rm Pl}   \Lambda_{\rm QCD}    }{ \sqrt{g_{\star}    }  } \simeq  10^{18} \left(   \frac{  \sigma_{0}}{ 10^{-31} {\rm \, cm}^2  }  \right).~~~
\ee
In \Eq{hubble-criterion} we have used \Eq{sigma-rel}, we have taken $T = \Lambda_{\rm QCD} = 200\mev$, we have assumed $n_q \sim \Lambda_{\rm QCD}^3$, and $g_{\star}(T=\Lambda_{\rm QCD}) = 61.75$ is the effective number of relativistic species just before the QCD phase transition.
Because \Eq{hubble-criterion} implies  $\Gamma/ H \gg 1$ for all cross sections that 
predict detectable CRUD, the general argument in \Sec{BBN} indicates that the contact-interaction regime is excluded for all DM masses below $\sim \few \mev$ (see \Eq{massbound}), independent of their other properties. 
 
We point out that comparing the rate to Hubble at $T\sim \Lambda_{\rm QCD}$ to assess thermalization is conservative because, for contact interactions, the ratio $\Gamma/H$ in fact {\it increases} for higher temperatures. A much stronger bound could be extracted by comparing these rates at temperatures of order the mass of the heavy particle which was integrated out to 
realize the non-renormalizable coupling $G_{\chi q}$ in \Eq{Lint}; however doing so requires an assumption about the early universe at temperatures above the QCD phase transition.
Furthermore, we note that for CRUD sized cross sections, 
 $\gamma \gamma \leftrightarrow \chi \bar \chi$ reactions through hadronic loops maintain equilibrium well
below the QCD phase transition. This rate is only suppressed by a loop factor 
relative to $\sigma_0$ in \Eq{hubble-criterion}, so the entropy transfers from the decoupling/annihilation
of SM species do not weaken the BBN constraint.

In \Fig{fig:money}, we plot the strongest bounds on the model of \Eq{LintN} of which we are aware. These include direct detection experiments operated at underground facilities \cite{Akerib:2016vxi, Cui:2017nnn, Aprile:2018dbl, Agnes:2018ves, Abdelhameed:2019hmk, Liu:2019kzq}, with upper limits from \cite{Davis:2017noy, Kavanagh:2017cru, Emken:2018run} where available and as described in \cite{Hooper:2018bfw} for the underground run of \cite{Liu:2019kzq}, which extends the low-mass reach of conventional detectors. (In attempting to use the techniques of \cite{Hooper:2018bfw} for the results in \cite{Collar:2018ydf}, we find that the optical depth at the best current limit is greater than unity, and thus interpretation of these results at all cross sections requires modeling of the velocity distribution as in \cite{Emken:2018run}.) We also show results from the XQC satellite experiment \cite{Wandelt:2000ad, Erickcek:2007jv}, displaying for completeness the possible effect of an efficiency factor as low as $0.2\%$ as suggested by \cite{Mahdawi:2018euy} with a dash-dotted line. Constraints from the population of Milky Way satellites \cite{Nadler:2019zrb} are applicable, as are results from the cooling of Supernova 1987A \cite{DeRocco:2019jti}, which we obtain by simply rescaling their dimensionless parameter $y$ by the dark matter mass. The neutrino floor is shown for recoils off of superfluid ${}^4$He~\cite{Battaglieri:2017aum}. Our results rule out the region above the blue dashed line from the necessity of a new mediator with mass $m_{\rm med} \leq 10 \mev$ to produce such large scattering cross sections, and the red shaded region from violating \Eqs{hubble-criterion}{massbound}.

We do not fill in the regions of \cite{Bringmann:2018cvk, Cappiello:2019qsw} because these studies did not incorporate 
energy dependence in the CR-DM cross section. CRUD requires a relativistic cross section which is incommensurate with the constraints from the other experiments shown in \Fig{fig:money}. A hypothetical interaction which does permit such a direct comparison by a completely energy-independent cross section, of which we are not aware, should take into account the other bounds presented here. If on the other hand the analyses in \cite{Bringmann:2018cvk, Cappiello:2019qsw} had considered a contact interaction 
with appropriate CR-energy dependence in the numerator it is likely that these 
exclusion regions would shift towards smaller cross-sections when translated into the $\sigma_0$-$m_\chi$
plane, but such an analysis is beyond the scope of this work. Furthermore, this shift would be mass-independent and would not evade the BBN exclusion region presented here.

\section{Light Mediators}
\label{mediators}

Previous studies of CRUD have assumed that the $\chi$-$ n$ scattering cross section is independent of energy, as discussed in \Sec{intro}. In \Sec{contact} we refined this ansatz by considering a physically well-defined contact-interaction scenario with an effective operator, and we included energy dependence in the cross section. However, 
realizing the cross sections needed for observable CRUD leads to an irreducible tension in the contact-interaction scenario: the existence of a non-renormalizable contact interaction imposes a lower bound on the mediator mass $m_{\rm med} \gtrsim E_{\rm CM} \sim \cO(\gev)$, where $E_{\rm CM}$ is obtained in \Eq{ecr-gev}, whereas realizing the large cross sections 
$\sigma_{0} \gtrsim 10^{-31}$ cm$^2$ necessary for detectable rates imposes an upper bound on the mediator particle mass. Indeed, if  $g_{\chi, N}$ is the mediator
 coupling to DM and nucleons respectively, the cross section in this regime is roughly 
\be
\label{mediator-requirement}
\sigma_{0} 
\sim  10^{-31}\cm^2  \, g_\chi^2 g_N^2  \pL \frac{\mu_{\chi N}}\mev \pR^2 \pL \frac{0.25\gev}{m_{\rm med}} \pR^4
\ee
so it is generically difficult to realize a large cross section without a light mediator. 
This tension suggests that a contact interaction may inadequately describe the scattering processes of interest. In this section, we therefore address whether a light mediator
can evade the argument put forth in \Sec{BBN}.

For concreteness, consider the renormalizable Lagrangian 
\be
\label{Leff-q}
{\cal L}_{\rm int} =  V_\mu\left(     g_\chi   \bar \chi \gamma^\mu \chi + g_q \bar q \gamma^\mu q \right)~,
\ee
where $V$ is a mediator particle with couplings $g_{\chi, q}$ and $g_N = \sum_q g_q$ is the resulting nucleon
coupling when the sum is over valence quarks.  
Here we take $m_V \ll E_{\rm CM}, \Lambda_{\rm QCD}$, so the mass
can be neglected in our estimates. 
For simplicity, $V$ is chosen to be a vector, but other choices yield similar 
conclusions up to order-one factors.  

Since realizing an appreciable CRUD effect requires a large cross section $\sigma_{\rm CR} \gtrsim 10^{-31} $ cm$^2$ in CR-DM interactions, this 
imposes a minimum requirement on the $V$ couplings. Approximating this cross section as
\be \label{light-med-sXN}
\sigma_{\rm CR} \simeq  \frac{g_\chi^2 g_q^2}{E_{\rm CM}^2} \simeq \frac{g_\chi^2 g_q^2}{ m_\chi E_{\rm CR} },
\ee 
 where the momentum transfer is taken to be $Q \simeq E_{\rm CM}$.
 Adopting the representative value $E_{\rm CR} \sim$ GeV where the CR spectrum approximately
peaks \cite{Patrignani:2016xqp}, we require the effective coupling to satisfy
\be
\label{crud-requirement-mediator}
\sqrt{ g_\chi g_q} \gtrsim   2\times 10^{-2}  \, \left[ \left(        \frac{\sigma_{\chi N}  }{10^{-31} \, \rm cm^2} \right)
\left( \frac{m_\chi}{ \rm MeV}  \right) \left( \frac{E_{\rm CR}}{ \rm GeV}  \right) \right]^{1/4} \hspace{-0.4cm},~~~~~~
\ee
for observable CRUD rates in the light mediator limit. 

We can now can ask: what values of $g$ suffice to equilibrate the DM and SM sectors while anti-quarks
are still present in the thermal bath around $T \sim \Lambda_{\rm QCD}$? Ignoring the order-one
difference between quark and nucleon couplings, we estimate the $\bar q q \to \bar \chi \chi$ 
annihilation rate as
\be
 \Gamma_{\bar q q \to \bar \chi \chi}  =  n_q \langle \sigma v \rangle \sim T^3 \frac{g_\chi^2 g_q^2 }{T^2}.
\ee
Comparing to the Hubble rate $H \sim \sqrt{g_\star} T^2/M_{\rm Pl}$, we find that
DM equilibrates at the QCD phase transition unless
\be
\label{thermalization-criterion-g}
\sqrt{g_\chi g_q} \lesssim 
 \left(   \sqrt{g_\star}  \frac{ \Lambda_{\rm QCD}}{ M_{\rm Pl}}     \right)^{1/4} \simeq 2 \times 10^{-5},
\ee
which is many orders of magnitude smaller than the CRUD requirement in
\Eq{crud-requirement-mediator}. It is therefore clear that thermalization is achieved even in the light (or massless)
mediator limit. Because CRUD and early-universe thermalization both take place
in the relativistic regime, the features of this argument are insensitive to the nature of the 
mediator or its Lorentz structure; any variations along these lines introduce at most order-one
differences. 

We finally point out that, in contrast with the heavy mediator regime in \Sec{contact}, this scenario is potentially {\it more} constrained by the BBN bound because the light mediator will also thermalize.
If such a population survives until BBN it will contribute additionally to $N_{\rm eff}$. In \Fig{fig:money} we show the region where the scattering cross section given by \Eq{mediator-requirement} is so large that it requires a mediator of mass less than 10 MeV. We do not impose any additional constraints on the mediator, although these considerations can be powerful \cite{Knapen:2017xzo, Digman:2019wdm}.

\section{Other Interaction Types}
\label{other}
We have addressed contact interactions in \Sec{contact} and
long range interactions in \Sec{mediators}. One might wonder whether operators with different Lorentz 
structure, higher dimension operators, or intermediate mass mediators could evade thermalization around $T=\Lambda_{\rm QCD}$.
Here we argue against these possibilities

\begin{itemize}
\item {\bf Vary Lorentz  Structure:} Since both CRUD and early universe thermalization
occur in the highly relativistic regime, changing the 
Lorentz structure in \Eq{Lint} or \Eq{Leff-q} would only affect rates by order-one factors.
By contrast, the thermalization criteria in \Eq{hubble-criterion} and \Eq{thermalization-criterion-g}
are violated by many orders of magnitude. Thus, the argument in this work covers
such variations equally well.

\item {\bf Higher Dimension Operators: } Consider DM-SM interactions mediated by higher-dimension operators characterized by some scale $\Lambda$ and some dimension $n$. Dimensional analysis requires
\be
~~~~~~~~~\frac{1}{\Lambda^n} {\cal \hat O}_\chi {\cal \hat O}_{\rm SM}  \implies  \langle \sigma v\rangle  \propto \frac{1}{\Lambda^2} \left(  \frac{T}{\Lambda} \right)^{2n-2}  \!\!\! ,~
\ee
where ${\cal O}_{\chi}$ and ${\cal O}_{\rm SM}$ are operators of DM and SM fields. We see that for $n >2$ the $\chi$-SM interaction rate
 $\Gamma/H$ would be even larger than in the 
simpler $n = 2$  regime studied in \Sec{contact}, so thermalization is even easier to achieve in this scenario for a fixed direct-detection cross section. 

\item {\bf Comparable Mediator Mass:} 
Thus far, we have excluded the $< 5$ MeV CRUD parameter space from BBN-only bounds on $N_{\rm eff}$. In \Sec{contact} and \Sec{mediators} we found that heavy-mediator (contact interaction) and
light-mediator scenarios, respectively, were excluded. What about an intermediate regime where the mediator is comparable in mass to the DM? In this case, the CR scattering will be suppressed compared to that in \Eq{light-med-sXN}, requiring even larger couplings. Furthermore, the mediator, which is also inevitably thermalized in the early universe, will contribute more to $\rho_{\rm tot}$ compared to \Sec{contact}. Both of these effects worsen the agreement with BBN observations.

\item {\bf DM With SM Gauge Charge: }
If $\chi$ is charged under the $SU(3)_c \times SU(2)_L  \times U(1)_Y$ gauge group of the SM, then it may possibly have a large interaction cross section due to SM gauge boson exchange. 
However, for $m_\chi \lesssim m_Z/2$, $SU(2)_L$ charged particles would contribute
unacceptably to the inferred number of active neutrino species from the invisible $Z$ width \cite{ALEPH:2005ab,Patrignani:2016xqp}.
 Furthermore, since CRUD requires $m_\chi \lesssim 100$ MeV, this excludes the possibility 
that DM could be a bound state of QCD charged particles, since such a state satisfies $m_\chi \gtrsim \Lambda_{\rm QCD} \sim 200$ MeV due to confinement, which
is outside the observable CRUD range.
Finally, if DM carries a QED ``millicharge," from \Eq{crud-requirement-mediator} CRUD requires
 $g_\chi/e  \gtrsim  10^{-2}  (m_\chi/100 \, \rm MeV)^{1/2}$, which is excluded over the 
 full keV-100 MeV range by accelerator searches for millicharged particles \cite{Prinz:1998ua,Davidson:2000hf}. 
 Thus, we conclude that there is no SM gauge interaction that can realize
an observable CRUD cross section $\gtrsim 10^{-31}$ cm$^2$ in the $\sim$ keV-100 MeV mass range; also
see \cite{Digman:2019wdm} for a discussion of realizing large DM interactions in theoretically consistent models.
\end{itemize}

\section{Laboratory Bounds}
\label{lab}
Thus far, we have found that BBN robustly excludes CRUD for $m_\chi
\lesssim $ few MeV regardless of whether it couples to quarks through a  dimension-6 contact-operator (\Sec{contact}), a
 light mediator (\Sec{mediators}), or some other exotic interaction (\Sec{other}). However, the CRUD parameter space of interest extends
 up to $m_\chi \sim$ 100 MeV, so for the $ m_\chi \sim$  few MeV$-$100 MeV range, we 
 turn to accelerator searches to constrain the DM and mediator that can realize this scenario.
 
 As in \Sec{mediators}, we consider the generic vector mediated interaction
 \be
\label{Leff-med}
{\cal L} \supset  V_\mu\left(     g_\chi   \bar \chi \gamma^\mu \chi + g_q \bar q \gamma^\mu q \right) + \frac{m_V^2}{2} V_\mu V^\mu~,
\ee
only now we require $ m_V \gtrsim 5$ MeV; by the same logic as in the earlier sections, 
the mediator is thermalized with the SM in the early universe so the same BBN bounds apply to $m_V$ even though 
$m_\chi$ satisfies \Eq{massbound}.

Generalizing \Eq{crud-requirement-mediator} for massive $V$, CRUD requires
\be
\label{crud-accelerator}
\sigma_{\rm CR} \sim 
\frac{g^2_q g_\chi^2  E_{\rm CM}^2 }{ {\rm max}(E_{\rm CM}^4, m_V^4 )   } \gtrsim 10^{-31} \ {\rm cm^2}.
\ee
Conservatively we take  $g_\chi = 4 \pi$ at the unitarity limit with $m_\chi = 10$ MeV and  $E_{\rm CR} =$ GeV,  so
 $E_{\rm CM} \sim 100$ MeV. Using these reference values, \Eq{crud-accelerator} requires  
 $g_q \gtrsim  0.01 \pL m_V / {\rm GeV}\pR^2$
for  $m_V \gg E_{\rm CM}$ and 
$g_q \gtrsim  10^{-4},$
for $m_V \ll E_{\rm CM}$.
These couplings are subject to a variety of laboratory searches. In the following, we 
will take the weaker  $g_q \sim 10^{-4}$ as the benchmark coupling required to realize observable CRUD.

SM couplings to sub-GeV scalar or pseudoscalar mediators break electroweak symmetry, so they are generically suppressed by factors of $\sim m_{u,d}/v \sim 10^{-5}$, where $v = 246$ GeV
is the Higgs vacuum value. Thus, 
it is difficult to satisfy $g_q \gtrsim 10^{-4}$ for a light (pseudo) scalar coupled to first generation quarks as required in \Eq{crud-accelerator} -- see \cite{Krnjaic:2015mbs} for further discussion.
 For this reason, we will continue to focus our discussion on a light vector.

Light vector mediators that couple to anomalous SM currents are generically very strongly constrained \cite{Kahn:2016vjr,Dror:2017ehi}, so any model with $m_V \lesssim$ GeV 
that hopes to realize $g_q \gtrsim 10^{-4}$, must identify $V$ with either a kinetically mixed ``dark photon" or the gauge boson of
 an anomaly-free $U(1)$ group that gauges a subset of SM quantum numbers. In the latter case, SM quarks carry charge under gauged
$B-L$ or $B-3L_i$ where $B$ and $L$ are  baryon and lepton number respectively and  $L_i$ is a lepton family number.

However, for $m_V \lesssim$ GeV,
in all of  these scenarios, the ``worst-case" reference coupling of $g_q \sim 10^{-4}$ is close to existing limits regardless of how $\chi$ decays \cite{Bauer:2018onh}. Thus, the bulk of the CRUD parameter space is likely ruled out by existing experiments and the high-mass lower boundary of CRUD parameter space is accessible at existing and proposed experiments  including ATLAS and CMS  \cite{Curtin:2014cca}, LSND \cite{deNiverville:2011it},  LHCb \cite{Ilten:2015hya}, Belle II \cite{DePietro:2018sgg}, MiniBooNE \cite{Aguilar-Arevalo:2018wea},  BDX \cite{Battaglieri:2016ggd}, SHiP \cite{Alekhin:2015byh}, NA62 \cite{CortinaGil:2019nuo}, NA64 \cite{NA64:2019imj}, LDMX \cite{Akesson:2018vlm,Berlin:2018bsc}, and HPS \cite{Celentano:2014wya} (see \cite{Battaglieri:2017aum} for a survey). Given the order-of-magnitude estimates presented here, dedicated studies are necessary to properly 
evaluate the possibility of complementarity between accelerators and direct detection experiments
for CRUD scenarios in the $m_\chi \sim\few\mev - \, 100 \mev$ range; such efforts are beyond the scope of the present work.

\section{Conclusions}
\label{conclusion}
In this paper we have identified a robust, nearly model-independent bound on scenarios in which cosmic rays 
upscatter light $<$ GeV dark matter. Since CRUD requires very large {\it relativistic} cross sections in order for
cosmic rays to upscatter halo DM particles, by crossing symmetry there is a comparably large cross section for 
DM to be thermalized in  the early universe through $\bar qq \to \bar \chi \chi$ annihilation with a rate 
predicted for a given CRUD cross section. For cross sections sufficiently large to enable CRUD 
detection $\sigma_{0} \gtrsim 10^{-31}$ cm$^2$, the DM particle is thermalized  with the SM at early times and
can contribute non-negligibly to the radiation density at BBN, thereby spoiling the successful prediction
of light-element yields in the SM.

For completeness, we note that there is a fundamental distinction between CRUD
and other large cross-section scenarios proposed in recent years. For instance, keV-MeV scale {\it freeze-in} DM
through an ultra-light mediator \cite{Chu:2011be, Dvorkin:2019zdi} evades the argument presented here. Namely, the direct-detection cross-section
for this freeze-in model can be
roughly in the same large CRUD range, but, unlike the CRUD scenario, in which all scattering or annihilation
processes are 
(quasi-)relativistic, the non-relativistic direct detection rate scales as  $\sigma  \propto  v^{-4} \sim 10^{12}$ where
 $v \sim 10^{-3}$ is the typical DM velocity. This scenario is safe from 
 thermalization in the early universe because the DM production rate
 is relativistic and scales as $\sigma \propto T^{-2}$, which can be orders of magnitude smaller at earlier times.
 
 The key observation in this paper is that, due to the relativistic kinematics of CR-DM scattering, the
 cross section for this process cannot be parametrically separated from the related DM production 
 cross section in the early universe. Thus, if the CRUD cross section is sufficiently large to enable terrestrial detection,
  then DM thermalization is a generic consequence of this scenario. $\Delta N_{\rm eff}$ bounds exclude 
  all such thermalized DM candidates with masses below a few MeV, which covers much of the favored $m_\chi \sim$ keV-100 MeV mass
  range over which CRUD can yield observable rates. For the remaining $\few\mev - \, 100 \mev$ window, we expect that accelerator
  searches for light DM and associated mediators place strong bounds on this scenario, but our crude estimates here 
  suggests that some parameter space may remain viable in this range. A careful, dedicated study is necessary to properly
  answer this question. However, if parameter space does remain viable, it will be within the reach of several
planned and proposed collider and fixed-target experiments, thereby offering complementary evidence in the 
event of a CRUD signal. 

Although our analysis here has emphasized the BBN bounds on CRUD, we note that these bounds 
also apply to a related scenario in which DM is relativistically produced in inelastic CR collisions and subsequently 
scatters in terrestrial detectors \cite{Alvey:2019zaa}. As in
the CRUD scenario, observable rates of DM production in these collisions require low $\sim$ keV-MeV DM masses and large, relativistic cross sections
$\gtrsim 10^{-33}$ cm$^2$, so an observable signal rate implies early-universe DM-SM thermalization and the corresponding BBN bound shown in Fig \ref{fig:money}.

\begin{acknowledgments}  
We would like to thank John Beacom, Christopher Cappiello, Alex Drlica-Wagner, Rouven Essig, Vera Gluscevic, and Maxim Pospelov for helpful conversations.
 This manuscript has been authored by Fermi Research Alliance, LLC under Contract No. DE-AC02-07CH11359 with the U.S. Department of Energy, Office of Science, Office of High Energy Physics. The United States Government retains and the publisher, by accepting the article for publication, acknowledges that the United States Government retains a non-exclusive, paid-up, irrevocable, world-wide license to publish or reproduce the published form of this manuscript, or allow others to do so, for United States Government purposes.
\end{acknowledgments}

\bibliography{crud}

\end{document}